\documentclass[11pt]{article}

\usepackage{amsmath}
\usepackage{amsthm,amssymb}
\usepackage[letterpaper,left=2.5cm,right=2.5cm,top=2cm,bottom=2cm]{geometry}

\usepackage{graphicx} \graphicspath{{figs/}}
\usepackage{subfigure}
\usepackage{psfrag}

\newcommand{\Ref}[1]{(\ref{#1})}

\newcommand{\diff}[2]{\frac{\partial #1}{\partial #2}}

\begin{document}
\title{On the location of the surface-attached globule phase in
  collapsing polymers.}
\author{A.\ L.\ Owczarek$^1$, A.\ Rechnitzer$^2$, J.\ Krawczyk$^1$ and T. Prellberg$^3$\\
  \footnotesize
  \begin{minipage}{13cm}
    $^1$ Department of Mathematics and Statistics,\\
    The University of Melbourne, Victoria~3010, Australia.\\
    \texttt{a.owczarek@ms.unimelb.edu.au,j.krawczyk@ms.unimelb.edu.au}\\[1ex]    
    $^2$ Department of Mathematics\\
    University of British Columbia, BC, V6T-1Z2, Canada\\
    \texttt{andrewr@math.ubc.ca}\\[1ex]    
$^3$ School of Mathematical Sciences\\
Queen Mary, University of London\\
Mile End Road, London E1 4NS, UK\\
\texttt{t.prellberg@qmul.ac.uk}
  \end{minipage}
}

\maketitle  

\begin{abstract}
  We investigate the existence and location of the \emph{surface}
  phase known as the ``Surface-Attached Globule" (SAG) conjectured
  previously to exist in lattice models of three-dimensional polymers
  when they are attached to a wall that has a short range potential.
  The bulk phase, where the attractive intra-polymer interactions are
  strong enough to cause a collapse of the polymer into a liquid-like
  globule and the wall either has weak attractive or repulsive
  interactions, is usually denoted \emph{Desorbed-Collapsed} or DC.
  Recently this DC phase was conjectured to harbour two surface phases
  separated by a boundary where the bulk free energy is analytic while
  the surface free energy is singular. The surface phase for more
  attractive values of the wall interaction is the SAG phase. We
  discuss more fully the properties of this proposed surface phase and
  provide Monte Carlo evidence for self-avoiding walks up to length
  256 that this surface phase most likely \emph{does} exist.
  Importantly, we discuss alternatives for the surface phase boundary.
  In particular, we conclude that this boundary may lie along the zero
  wall interaction line and the bulk phase boundaries rather than any
  new phase boundary curve.
 \end{abstract}

 The phase transitions of a single isolated polymer in solution
 continue to attract attention as single polymers are fundamental
 components in more complicated modelling scenarios and these
 transitions are not yet fully understood. The collapse transition
 \cite{gennes1979a-a} mediated by the intra-polymer attractive
 interactions and the adsorption transition \cite{debell1993a-a} when
 a polymer is attached to a sticky wall, are two of the key transitions
 that have been well studied. The situation when both transitions can
 occur in the same system has been studied less intensely due to the
 difficulty of numerical work such as Monte Carlo simulations when two
 independent parameters compete.  However, with the advent of more
 powerful computers and sophisticated algorithms this situation has
 received some attention.
 
 The standard lattice model for polymers is the self-avoiding walk
 (SAW) with the sites of the walk called monomers. The intra-polymer
 attraction is modelled by a potential energy $\varepsilon_p$
 associated with monomers that are nearest-neighbours on the lattice.
 These instances will be referred to as (nearest-neighbour)
 \emph{contacts}. To consider adsorption a wall is introduced, so that
 the polymer is restricted to one side of the wall, or may visit the
 wall, with one end of the polymer (SAW) attached to the wall.  Any
 monomer which visits the wall, other that the one fixed on the wall,
 is given a potential energy $\varepsilon_w$. These monomers will be
 referred to as \emph{visits}. Various different phases are
 conjectured to exist at different values of the ratio of the two
 energies and the temperature (see below). We shall restrict our
 discussion to three dimensions, as this is where more complex
 behaviour can occur (rather than two dimensions, where more is known
 due to exactly solved models and conformal field theory).
 
 Consider the simple cubic lattice and the half space $z\geq 0$.
 Consider an $n$-step self-avoiding walk $\varphi$ with one end fixed
 at the origin. We define two Boltzmann weights
 $\omega_p=e^{-\varepsilon_p/k_B T}$ and
 $\omega_w=e^{-\varepsilon_w/k_B T}$ where $k_B$ is Boltzmann's
 constant and $T$ is the temperature. Therefore, when $\omega_w > 1$,
 $\omega_w = 1$ and $\omega_w < 1$ the interaction of the walk with
 the wall is attractive, neutral, and repulsive, respectively. When $\omega_p >
 1$, $\omega_p = 1$ and $\omega_p < 1$ the intra-polymer interaction is
 attractive, neutral, and repulsive, respectively.

\begin{figure}[ht!]
  \centering
  \includegraphics[scale=1.0]{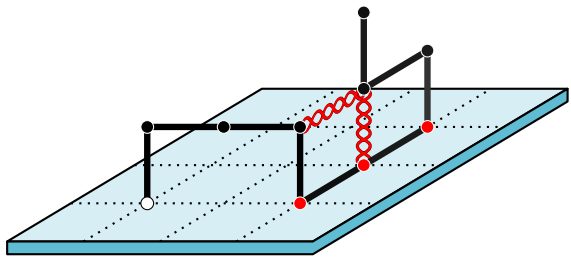}
  \caption{A 9-step SAW ($n=9$) attached to a wall with three visits ($m_w=3$)
    to the wall and two nearest neighbour contacts ($m_w=2$, shown as double helices).}
  \label{config}
\end{figure}

The partition
$Z_n(\omega_p,\omega_w)$ of the model (see Figure~\ref{config}) is defined as
\begin{equation}
Z_n(\omega_p,\omega_w)=\sum_{\varphi \in \Omega_n}
\omega_p^{m_p(\varphi)} \omega_w^{m_w(\varphi)}\; ,
\end{equation}
where $\Omega_n$ is the set of all $n$-step SAW restricted to the half
space $z\geq 0$, $m_p(\varphi)$ is the number of contacts in the walk
$\varphi$, and $m_w(\varphi)$ is the number of wall visits of the walk
$\varphi$. The \emph{bulk} thermodynamic (reduced) free energy is
given by the limit
\begin{equation}
f_b(\omega_p,\omega_w) = - \lim_{n\rightarrow\infty} 
\frac{1}{n}\log\big(Z_n(\omega_p,\omega_w) \big)\; .
\end{equation}

The \emph{bulk} phase diagram is determined by the analytic structure
of $f_b(\omega_p,\omega_w)$. This phase structure was studied in a
series of papers \cite{vrbova1996a-a,vrbova1998a-a,vrbova1999a-a}
where a schematic phase diagram was proposed (see Figure~\ref{fig
  phase diagram}). To discuss this diagram it is worth considering two
order parameters.  Firstly let us define the ``internal density''
$\bar{\rho}_p$ as
\begin{equation}
\bar{\rho}_p = \lim_{n\rightarrow \infty} \rho_p(n) =
\lim_{n\rightarrow \infty} \frac{n}{R_g^3(n)}\; ,
\end{equation}
where $R_g(n)$ is the radius of gyration. Secondly we define the
density of wall visits $\bar{\rho}_w$
as
\begin{equation}
\bar{\rho}_w = \lim_{n\rightarrow \infty} \rho_w(n) = \lim_{n\rightarrow
  \infty} \frac{\langle m_w\rangle}{n}\; ,
\end{equation}
where $\langle m_w\rangle$ is the ensemble average of the
number of visits to the wall. 
\begin{figure}[ht!]
  \centering
  \includegraphics[scale=0.5]{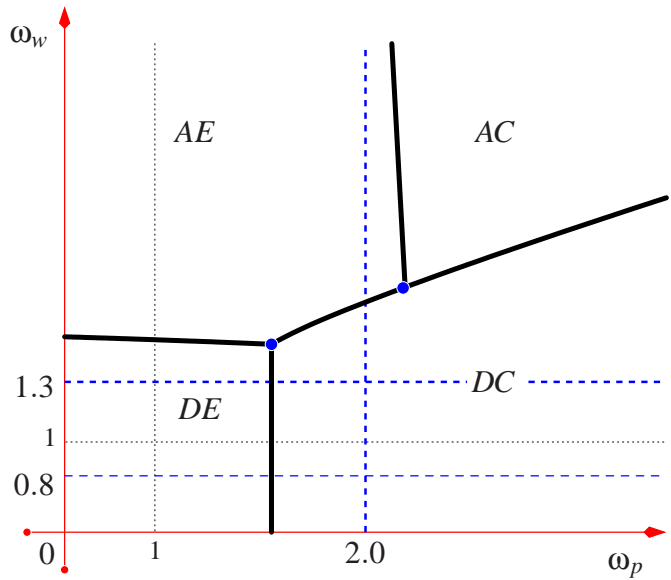}
  \caption{Schematic bulk phase diagram for collapsing adsorbing walks
    in three dimensions showing the four phases Desorbed-Extended
    (DE), Desorbed-Collapsed (DC), Adsorbed-Extended (AE) and
Adsorbed-Collapsed (AC). Also shown are the dashed lines where
simulations were performed.}
  \label{fig phase diagram}
\end{figure}

For any ratio of wall to intra-polymer energies at high temperatures,
the polymer is in the excluded volume state and is entropically
repulsed from the wall. It is expected that $ R_g(n) \sim n^{\nu}$ as
$n\rightarrow\infty$, where the three-dimensional excluded volume
value has been estimated as $\nu=\nu_3 \approx 0.5874(2)$
\cite{prellberg2001b-a}. The average number of visits is expected to
behave as $\langle m_w\rangle =o(n)$ (in fact, it is expected to be
bounded): this has been numerically verified.  Hence for
$\omega_w=\omega_p=1$ we have $\bar{\rho}_w=\bar{\rho}_p=0$. Since
$\bar{\rho}_p=0$ the polymer is referred to as ``Extended'' and since
$\bar{\rho}_w=0$ the polymer is referred to as ``Desorbed''. 

Now let us consider fixed $\omega_w=1$ and varying $\omega_p$.  That
is, we consider the collapse transition of a polymer fixed to a
non-interacting wall as the intra-polymer attraction is increased. As
the temperature is lowered (increasing the effect of the attraction)
the collapse transition is understood to occur at the $\theta$-point,
where for high temperatures $\bar{\rho}_p=0$ (the Extended phase),
while for low temperatures $\nu = 1/3$ and $\bar{\rho}_p>0$ (the
Collapsed phase): this is expected to be a second order phase
transition \cite{gennes1979a-a}.  In both two and three dimensions
this transition has been extensively studied (see for example
\cite{vanderzande1998a-a} and references therein).  

On the other hand, if we consider fixed $\omega_p=1$ as the
temperature is lowered ($\omega_w$ is increased), the wall density is
expected to change from $\bar{\rho}_w=0$ (Desorbed phase) for high
temperatures to $\bar{\rho}_w>0$ (Adsorbed phase) for low
temperatures.  This adsorption transition is also expected to be
second order \cite{debell1993a-a} and also has been well studied with
this description being numerically well verified (again also see
\cite{vanderzande1998a-a}). Note that for low temperatures the polymer
is in a two-dimensional excluded volume state, where $\nu=\nu_2=3/4$
\cite{nienhuis1982a-a}. We know that a two-dimensional polymer can
also admit a collapse transition, so one naturally can conjecture a
state where the polymer is both adsorbed with $\bar{\rho}_w>0$ and
two-dimensionally collapsed. A suitable two-dimensional order
parameter to describe this transition is
\begin{equation}
\bar{\sigma}_p = \lim_{n\rightarrow \infty} \sigma_p(n) =
\lim_{n\rightarrow \infty} \frac{n}{R_g^2(n)}\; .
\end{equation}

When considering the full problem where both $\omega_p$ and $\omega_w$
can vary Vrbov\'{a} and Whittington \cite{vrbova1996a-a,vrbova1998a-a}
conjectured, and \emph{numerically verified}, a phase diagram with
precisely these four phases: Desorbed-Extended (DE) phase with $\bar{\rho}_p =0$
and $\bar{\rho}_w=0$ for small $\omega_p$ and $\omega_w$, Desorbed-Collapsed (DC) phase with $\bar{\rho}_p> 0$ and
$\bar{\rho}_w=0$ for large $\omega_p$ and (relatively) small $\omega_w$,
Adsorbed-Extended (AE) with $\bar{\sigma}_p=0$ and $\bar{\rho}_w>0$
for (relatively) small $\omega_p$ and large $\omega_w$,
and Adsorbed-Collapsed (AC) with $\bar{\sigma}_p>0$ and
$\bar{\rho}_w>0$ for large  $\omega_p$ and $\omega_w$.

Let us return to the bulk Desorbed-Collapsed phase where
$\bar{\rho}_p> 0$ and $\bar{\rho}_w=0$. One may now ask: how does
$\bar{\rho}_w$ approach zero as the length is increased or, rather
equivalently, how does the average number of wall visits $\langle
m_w\rangle$ scale as $n$ is increased. It is believed (see below) that
in the Desorbed-Extended phase $\langle m_w\rangle =O(1)$, that is it
is bounded. One may be tempted to assume that this is also the case in
the Desorbed-Collapsed phase. However,
recent work \cite{singh2001a-a,rajesh2002a-a,mishra2003b-a} suggests
that two different asymptotic behaviours can occur inside the
latter.

To understand this, let us consider the
finite-sized \emph{extensive} (reduced) free energy
$F(n;\omega_p,\omega_w)= -\log(Z_n)$. From thermodynamics we expect that
\begin{equation}
F(n;\omega_p,\omega_w) = f_b(\omega_p,\omega_w) n +o(n)\; ,
\end{equation}
and as stated above it is the analytic structure of the bulk free
energy $f_b(\omega_p,\omega_w)$ that defines the phases of our model.
Drawing on the standard polymer scaling in \cite{owczarek1993b-:a}, the
behaviour of the finite-size free energy in the DE phase is given by
\begin{equation}
\label{eqn DE scaling}
F(n;\omega_p,\omega_w) = f_b(\omega_p,\omega_w) n + (\gamma_{1}-1)
\log(n)+ O(1)\; ,
\end{equation}
where $\gamma_{1}$ is an universal constant for the DE phase (an
exponent, in fact).
In the DC phase the scaling is expected to behave as
\begin{equation}
F(n;\omega_p,\omega_w) \sim f_b(\omega_p,\omega_w) n +f_s (\omega_p,\omega_w) n^{2/3},
\end{equation}
where we have a \emph{surface free energy} $f_s(\omega_p,\omega_w)$.
This behaviour arises as the polymer assumes a dense liquid-like drop
with a well defined \emph{surface} that has area $\propto n^{2/3}$. We
reiterate that this ``surface'' is \emph{not} the wall but rather the
surface of the liquid-like polymer drop. In the extended phase the
polymer does not have a well defined surface and so $f_s=0$ as in
equation~\Ref{eqn DE scaling}.

Importantly, the mean number of wall visits $\langle m_w \rangle$ can be calculated in the usual way as
\begin{equation}
\langle m_w \rangle = - \diff{F(n;\omega_p,\omega_w) }{\log(\omega_w)}\;.
\end{equation}
Hence
\begin{equation}
\langle m_w \rangle \sim -
\diff{f_b(\omega_p,\omega_w)}{\log(\omega_w)} n -
\diff{f_s(\omega_p,\omega_w) }{\log(\omega_w)} n^{2/3}\; .
\end{equation}
Since we assume that in both the Desorbed-Extended and
Desorbed-Collapsed phases $\langle m_w\rangle =o(n)$, it is clear that
$f_b(\omega_p,\omega_w)$ should not be depend on $\omega_w$. Hence we
have 
\begin{equation}
\langle m_w \rangle \sim - \diff{f_s(\omega_p,\omega_w)
}{\log(\omega_w)} n^{2/3}\; . 
\end{equation}
In the Desorbed-Extended phase $f_s(\omega_p,\omega_w)=0$, which
derives (with  some other weak assumptions) the $\langle m_w\rangle
=O(1)$ result for that phase. The question that then arises is the
behaviour of $f_s(\omega_p,\omega_w)$ for the Desorbed-Collapsed phase.

In \cite{singh2001a-a,rajesh2002a-a,mishra2003b-a} it was suggested
that the Desorbed-Collapsed phase accommodates two different
behaviours: one where $f_s(\omega_p,\omega_w)$ depends on $\omega_w$
and so $\langle m_w\rangle \sim n^{2/3}$, which was dubbed the
``Surface Attached Globule'' or SAG, and one in which
$f_s(\omega_p,\omega_w)$ is independent of $\omega_w$, so that $\langle
m_w\rangle =O(1)$ as in the Desorbed-Extended phase. We shall refer to this second
situation as the ``Fully Detached Globule''. Physically the SAG phase
can be pictured as the liquid drop partially wetting the wall, while
the FDG is the state where the wall is dry. (The Adsorbed-Collapsed
phase is the equivalent of a fully wet state.) This implies a surface
phase transition with a singularity in $f_s(\omega_p,\omega_w)$ as
$\omega_w$ is varied, whereas $f_b(\omega_p,\omega_w)$ is analytic at
the same point and no bulk transition occurs.

The evidence given in \cite{singh2001a-a,rajesh2002a-a,mishra2003b-a}
for the two `surface' states was based on the analysis of relatively
short exact enumeration data and some analysis of a directed walk
model in two dimensions. It is therefore important to test this
conjecture with data from longer walks. In this paper we have
simulated, using a recently developed Monte Carlo algorithm
\cite{prellberg2004a-a}, the model described above along various lines
in the parameter space for lengths up to 256. In a previous paper
\cite{krawczyk2005b-:a} we considered the bulk phase diagram and
various low temperature finite size features.

\begin{figure}[ht!]
  \centering \subfigure[]{
    \includegraphics[width=6cm]{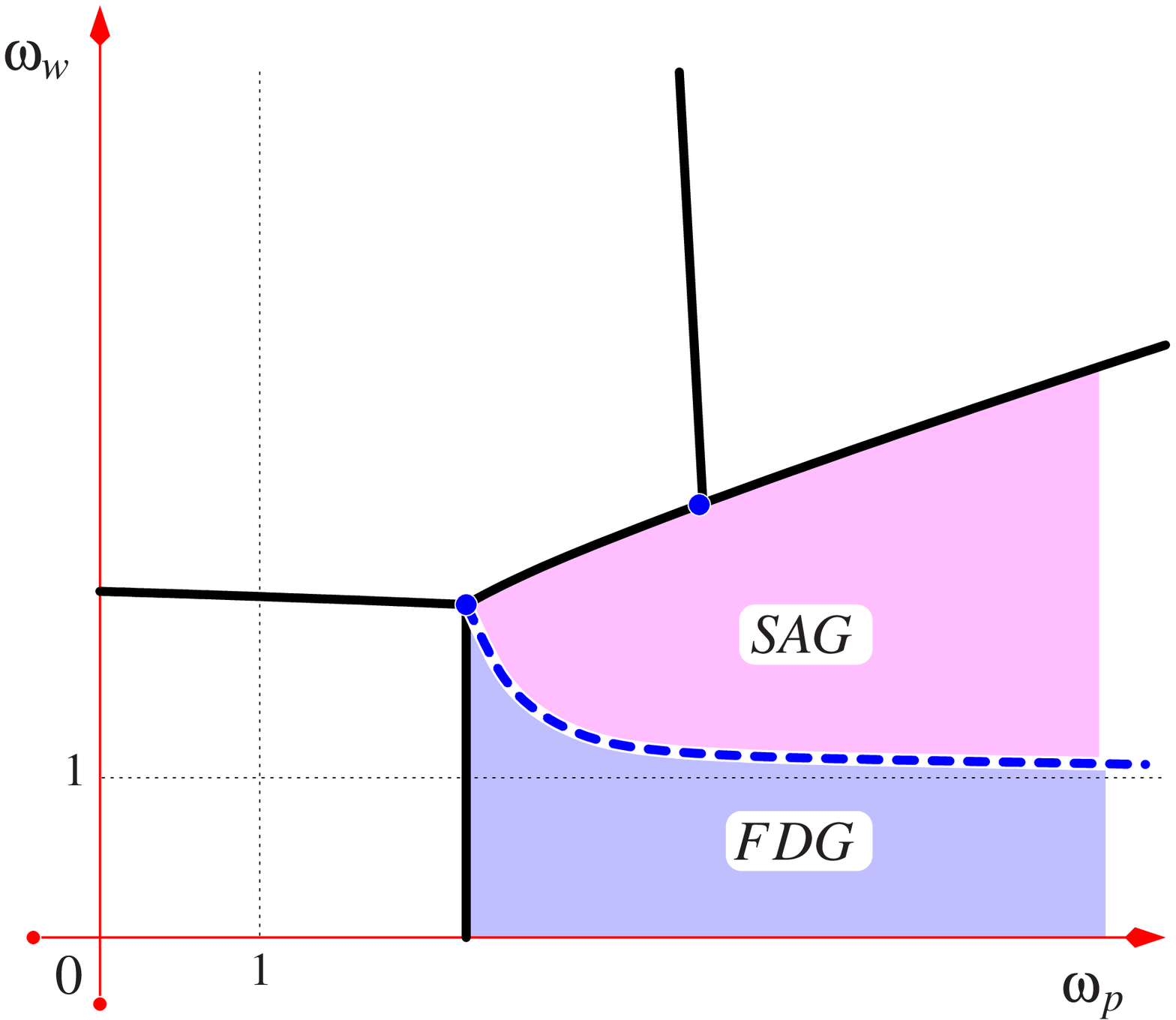}} \subfigure[]{
    \includegraphics[width=6cm]{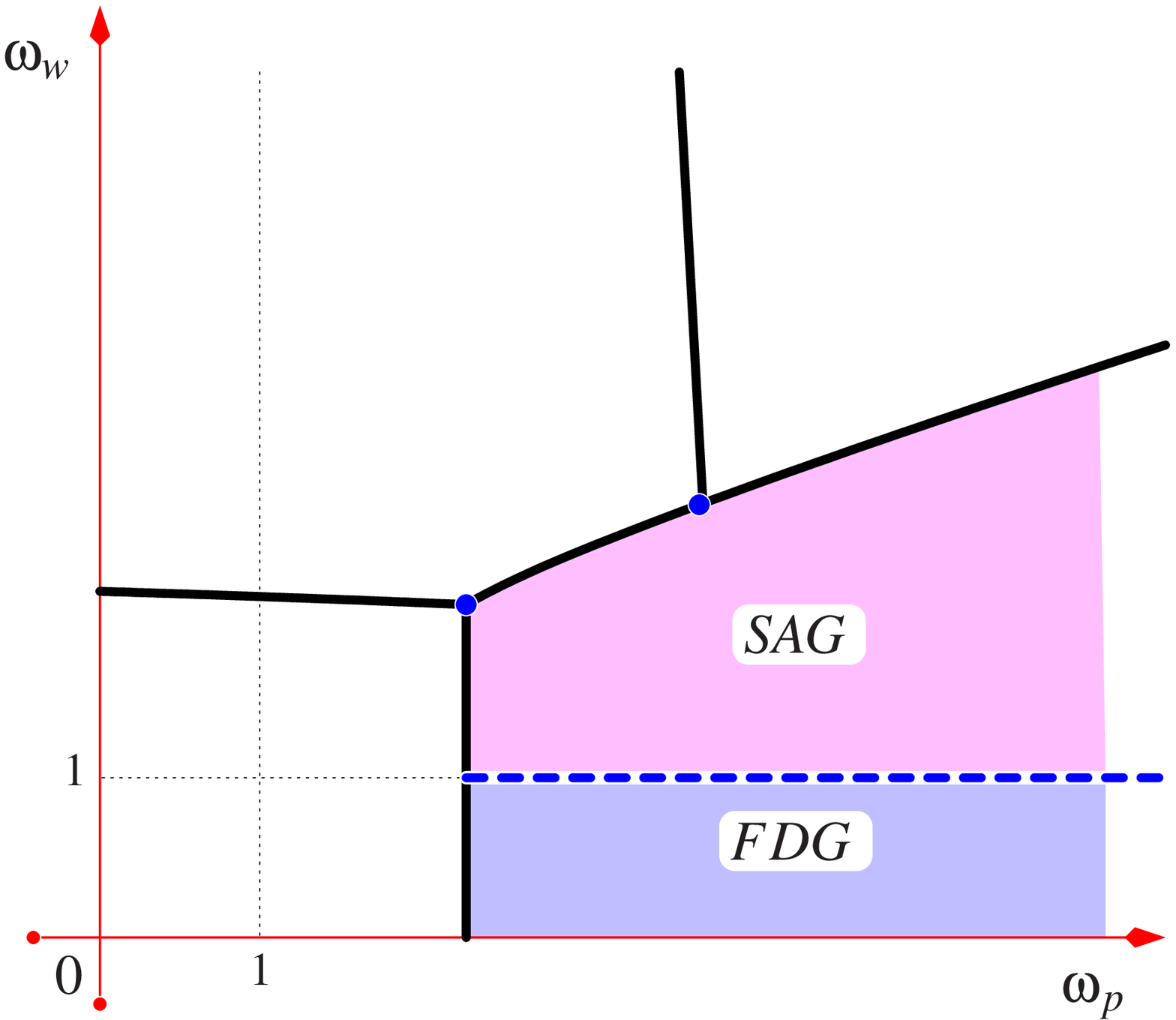}}
  \caption{Hypothetical boundary for the surface phase
    transitions. (a) conjectured in \cite{rajesh2002a-a}. (b) is an
    alternative that is suggested in this paper.}
  \label{fig hypotheses}
\end{figure}

In \cite{rajesh2002a-a} the boundary between the FDG and SAG was
conjectured (see Figure~\ref{fig hypotheses}~(a)). Using a zero
temperature argument \cite{rajesh2002a-a,krawczyk2005b-:a}, it is easy
to see that for infinite $\omega_p$ the transition from FDG to SAG occurs
at $\omega_w=1$. However, the question arises as to the form of the
phase boundary for finite $\omega_p$. In \cite{rajesh2002a-a} it was
conjectured that for finite $\omega_p$ (but large enough so that the
polymer is collapsed) the boundary occurs at values of $\omega_w$
greater than one. Hence, with small but finite attractive wall
potential the polymer is expected to still be in the FDG state.  On
the other hand, if there is no entropic penalty to be paid for the
globule to sit on the wall, this gives rise to an alternative
hypothesis for the FDG-SAG boundary as shown in Figure~\ref{fig
  hypotheses}~(b).
We note that the phase boundary is unlikely to lie in the region for
$\omega_w<1$; here the wall is repulsive and does not increase the
surface entropy of the polymer, so one may argue that typical
configurations have very few visits to the wall.  Hence one
conjectures that for $\omega_w<1$ the surface free energy
$f_s(\omega_p,\omega_w)$ is independent of $\omega_w$ and $\langle
m_w\rangle = O(1)$. The central question therefore is whether or not the
polymer-drop partially wets the wall as soon as the wall potential
becomes attractive. In the original hypothesis in Figure~\ref{fig
  hypotheses}~(a), a sufficiently attractive wall is required before
the polymer-drop partially wets the wall. The alternative proposed in
Figure~\ref{fig hypotheses}~(b) conjectures that any attractive
potential at the wall will induce the polymer-drop to partially wet the wall.

So as to delineate the SAG-FDG phase boundary and to test for the
existence of the SAG phase, we have studied three lines in the phase
diagram (see Figure~\ref{fig phase diagram}) . One line at
$\omega_p=2.0$ was chosen since the $\theta$ point is expected to be
around $\omega_p=1.5$ at the lengths considered here, and so by fixing
$\omega_p=2.0$ and varying $\omega_w$ one explores the
Desorbed-Collapsed phase. We note that the position of the collapse
transition should not move as $\omega_w$ is varied
\cite{vrbova1998a-a}.  To further consider the difference between
attractive and repulsive wall interactions on the Desorbed-Collapsed
phase, we have also simulated along lines $\omega_w=0.8$ (repulsive
wall visits) and $\omega_w=1.3$ (attractive wall visits). The second
of these lines also allows us to search for the surface phase boundary
proposed in Figure~\ref{fig hypotheses}~(a).  Note that the position
of the adsorption transition for small $\omega_p$ moves little as
$\omega_p$ is varied.

In order to explore the Desorbed-Collapsed phase, we have used the FlatPERM Monte
Carlo algorithm \cite{prellberg2004a-a}. This algorithm estimates
the density of states directly and so allows us to compute the
partition function and related quantities for a wide range of
$\omega_p$ and $\omega_s$. The memory required to store the density of
states grows with the cube of the length and so we were restricted to
SAWS of maximum length 256. Each simulation sampled approximately
$10^9$ confirmations. To reduce errors we performed 25 independent
simulations and combined the resulting data.

Before we discuss the subtleties of the surface phenomena of the SAG
and FDG phases we first verify that our simulations were correctly
identifying the bulk phases. To do this we estimated the average
end-to-end distance of the polymer which has been established to scale
in the same manner as the (more computationally costly to estimate)
radius of gyration. Hence we could estimate the size exponent $\nu$ at
various points. This confirms earlier work \cite{vrbova1998a-a}. We
considered the point $(\omega_p,\omega_w)= (1,1)$, which is in the
Desorbed-Extended phase, and so should have $\nu\approx 0.59$: we find
an estimate 0.60(2).  Next, for the point $(\omega_p,\omega_w)= (1,2)$
which is expected to be in the Adsorbed-Extended phase and so should
have $\nu= 3/4$ we find an estimate 0.75(1). Similarly, for the points
$(\omega_p,\omega_w)= (2,0.8)$ and $(2,1.3)$ which are in the
Desorbed-Collapsed phase (and we identify below to be in the FDG
surface and SAG surface phases respectively) and so should have $\nu=
1/3$ we find equal estimates 0.33(2).

\begin{figure}[ht!]
  \centering
  \psfrag{omegab}{$\omega_p$}
  \includegraphics[width=9cm]{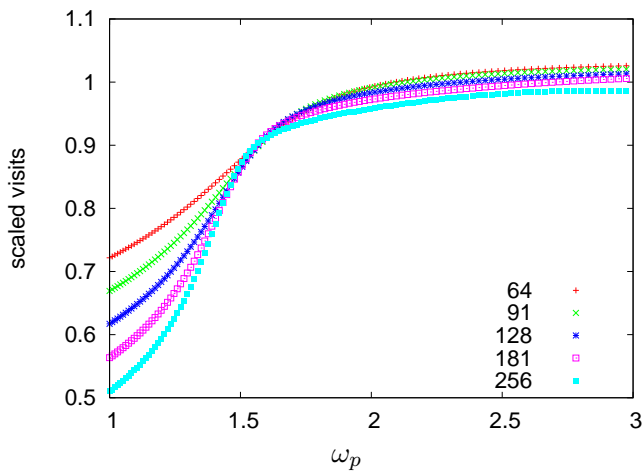}
  \caption{The mean number of visits divided by $n^{2/3}$ against
    $\omega_p>1$ for fixed $\omega_w=1.3$.}
  \label{fig sag exists}
\end{figure}

We begin our discussion of the surface phenomena by verifying the
existence of the SAG phase.  Let us consider the lines $\omega_w=1.3$
(attractive wall visits) and $\omega_w=0.8$ (repulsive wall visits).
In Figure~\ref{fig sag exists}, the mean number of visits
$m_w(n;\omega_p,1.3)$ divided by $n^{2/3}$ is plotted against
$\omega_p$ for $1.0\leq \omega_p\leq 3.0$ and $n=64$, $91$, $128$,
$181$, and $256$. The quantity $m_w(n)/n^{2/3}$ would be expected to
converge to a finite (non-zero) value only in the SAG phase. While
there are clearly some corrections to scaling still evident at these
lengths, $m_w/n^{2/3}$ seems to be convergent to a non-zero value for
$\omega_p \gtrsim 1.5$, which is the rough location of the collapse
transition at this length (the thermodynamic location has been
estimated to be near $1.3$).  We have also checked that for $\omega_p
<< 1.5$ (that is, in the DE bulk phase) $m_w$ converges to a finite
value.  Hence the (finite-size location) bulk phase transition at
$1.5$ signifies a change in the length scaling of $m_w(n)$ from $n^0$
for small $\omega_p$ (DE bulk phase) to $n^{2/3}$ for large $\omega_p$
(DC bulk phase).

\begin{figure}[ht!]
  \centering
  \psfrag{omegab}{$\omega_p$}
  \includegraphics[width=9cm]{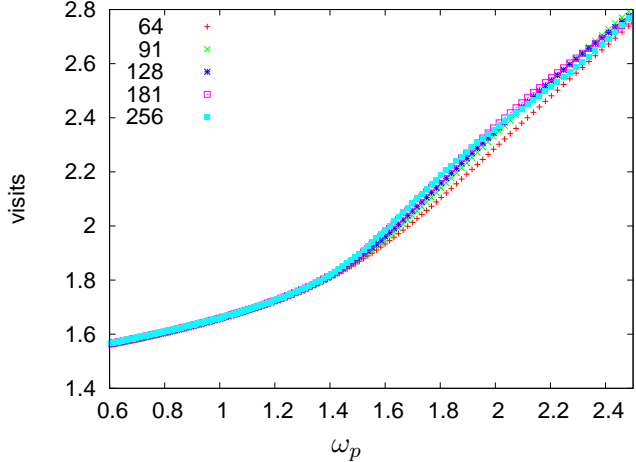}
  \caption{The mean number of visits against
    $\omega_p$ for fixed $\omega_w=0.8$.}
  \label{fig fdg exists}
\end{figure}

In Figure~\ref{fig fdg exists}, the raw mean number of visits
$m_w(n;\omega_p,0.8)$ is plotted against $\omega_p$ for $1.0\leq
\omega_p\leq 3.0$ and $n=64$, $91$, $128$, $181$, and $256$. As with
$\omega_w=1.3$ one expects that for $\omega_p < 1.5$ (DE bulk phase)
that $m_w(n)$ will converge to a finite value as $n$ is increased and
indeed this is the case. Additionally, the quantity $m_w(n)$ would be
expected to converge to a finite (non-zero) value in the FDG phase.
It appears that the collapse transition at around $\omega_p=1.5$ only
effects the value of $\lim_{n\rightarrow\infty}m_w(n)$ but not whether
it is finite or not.  There is no indication of a change of length
scaling when varying~$\omega_p$.

\begin{figure}[ht!]
  \centering
  \psfrag{omegab}{$\omega_p$}
  \subfigure[]{ \includegraphics[width=7.5cm]{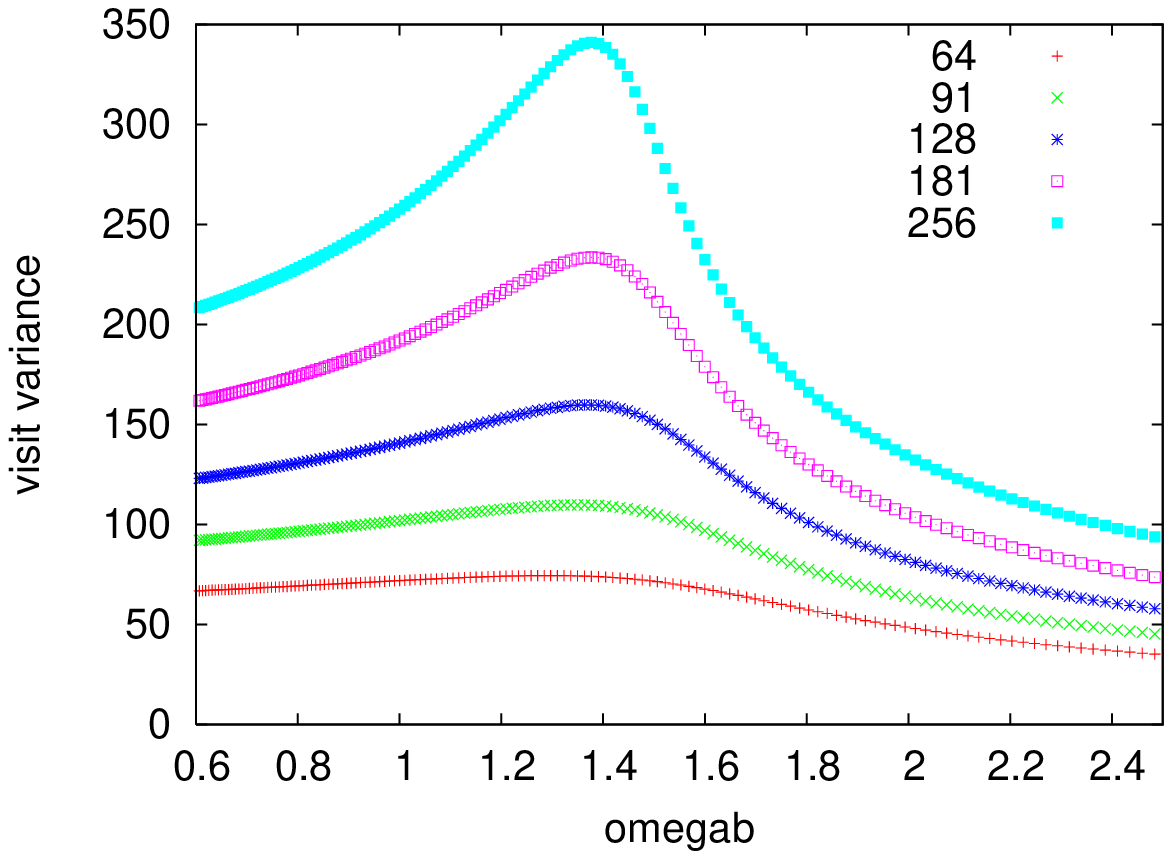}}
  \subfigure[]{ \includegraphics[width=7.5cm]{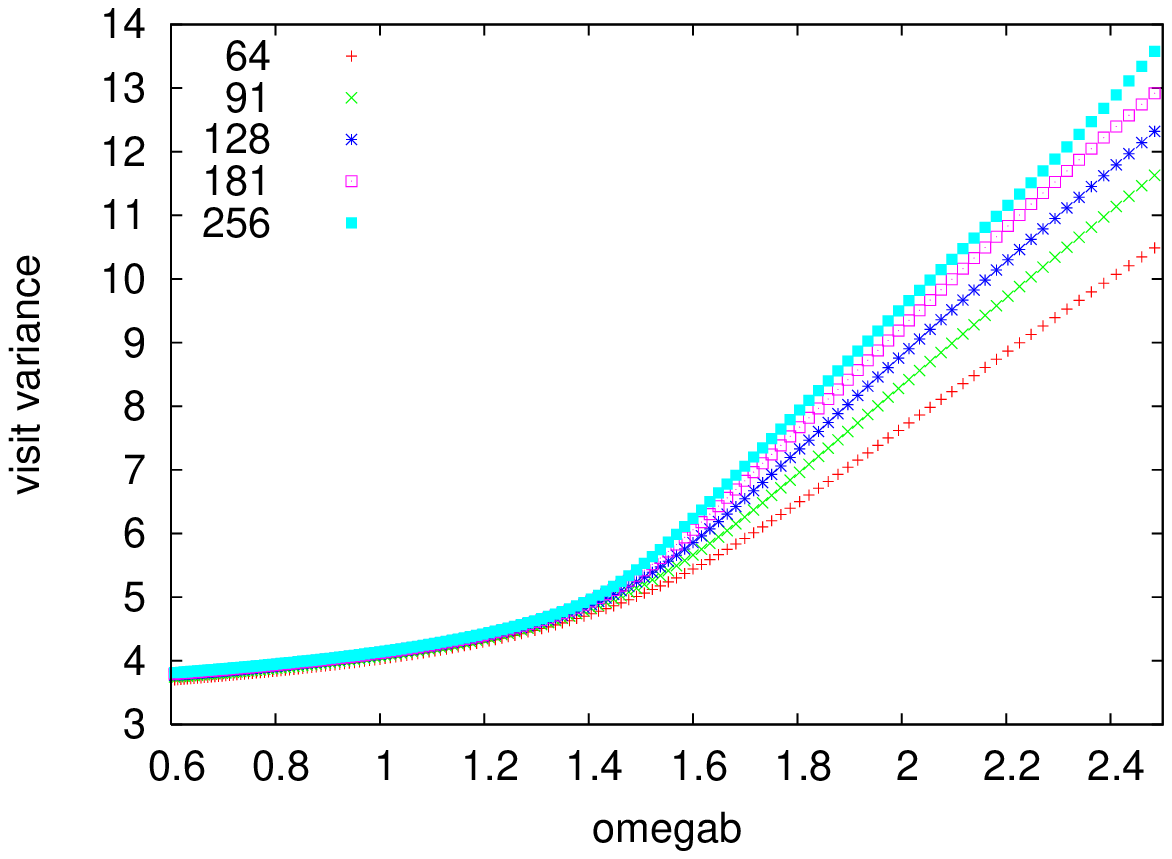}}
  \caption{Plot of the variance in the number of visits vs $\omega_p$
    for $\omega_w=1.3$~(a) and $\omega_w=0.8$~(b). This
    clearly shows a difference between SAG and FDG.}
  \label{fig comp sag fdg variance}
\end{figure}

In Figure~\ref{fig comp sag fdg variance}, the variance of the number
of visits is plotted against $\omega_p$ for $\omega_w=1.3$~(a) and
$\omega_w=0.8$~(b). In (a) a single peak in the variance is developing
 around $\omega_p=1.4$ as the length is increased, whereas in (b) no such peak
is developing. We conclude a wall visit transition is occurring at
around the same place the bulk contact transition occurs when
$\omega_w=1.3$ \emph{but not} when $\omega_w=0.8$.

\begin{figure}[pht!]
  \centering
  \psfrag{omegas}{$\omega_w$}
  \subfigure[]{ \includegraphics[width=7.5cm]{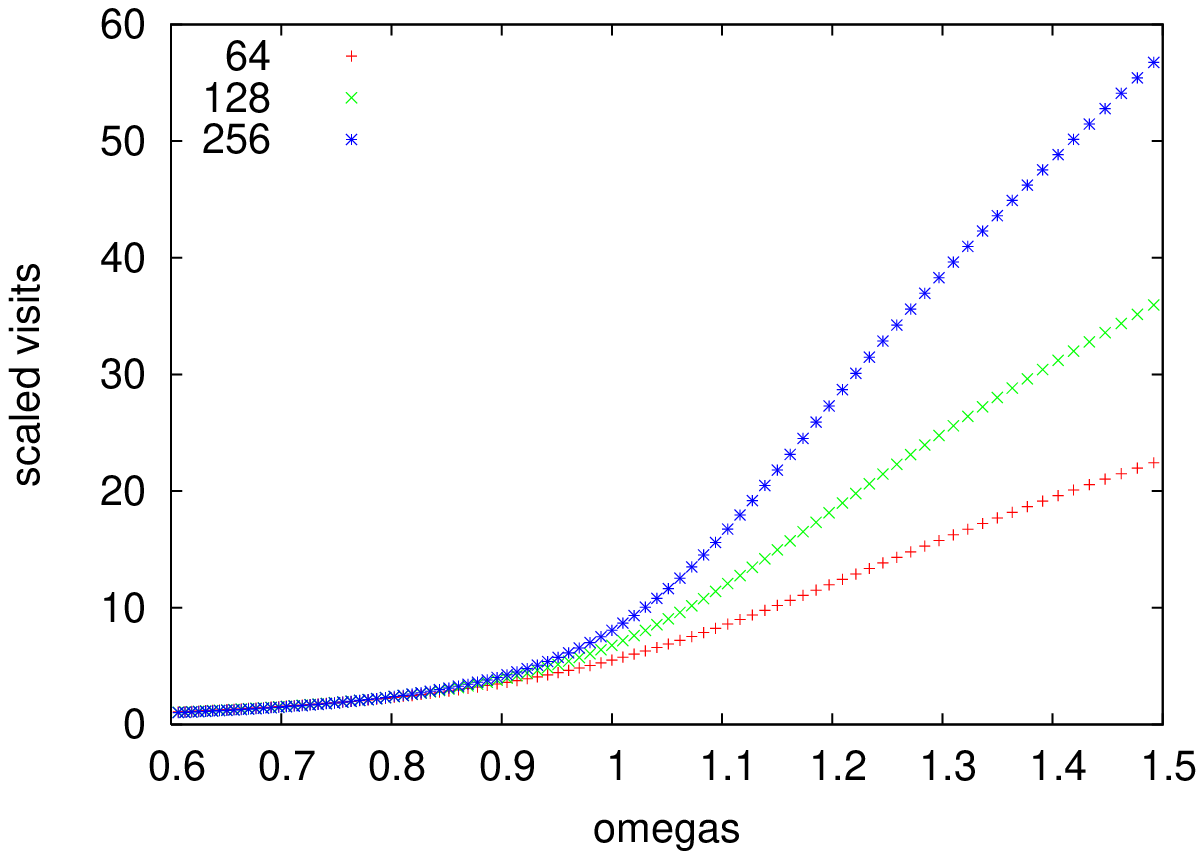}}
  \subfigure[]{ \includegraphics[width=7.5cm]{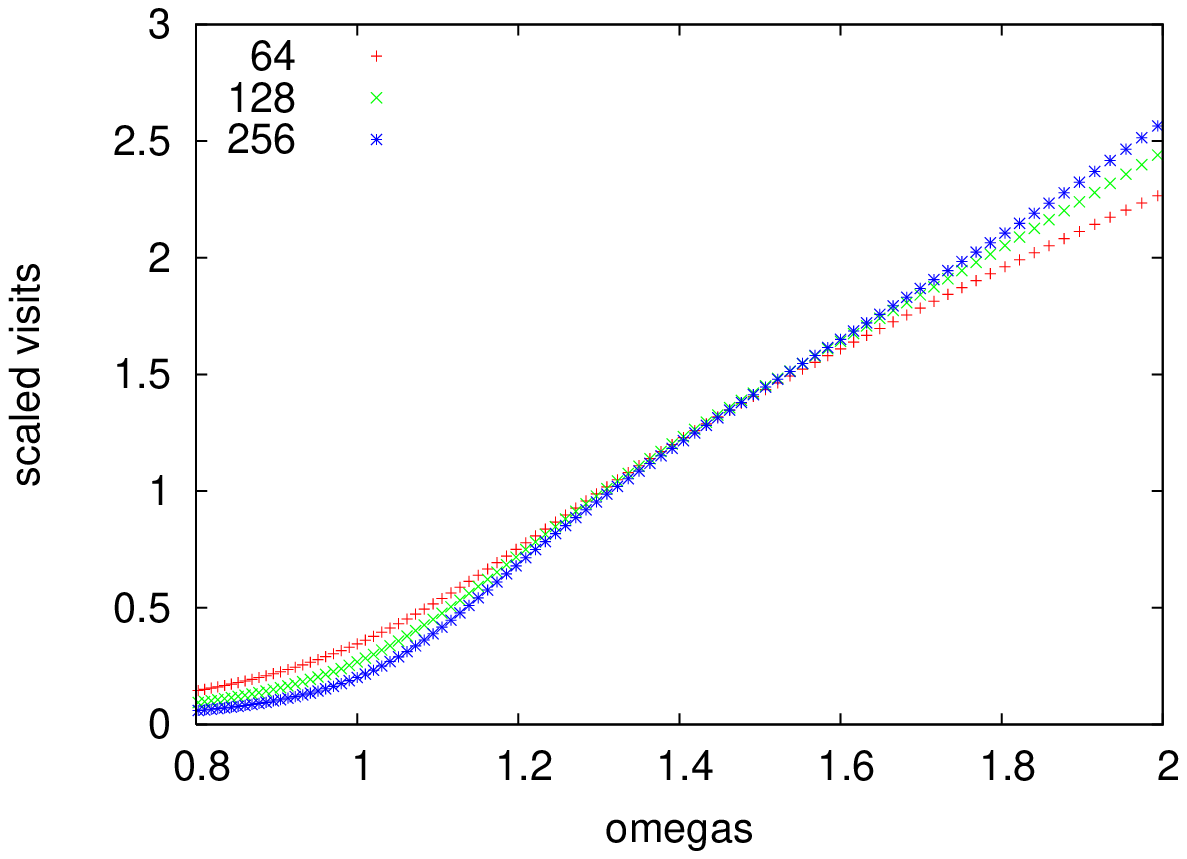}}
  \subfigure[]{ \includegraphics[width=7.5cm]{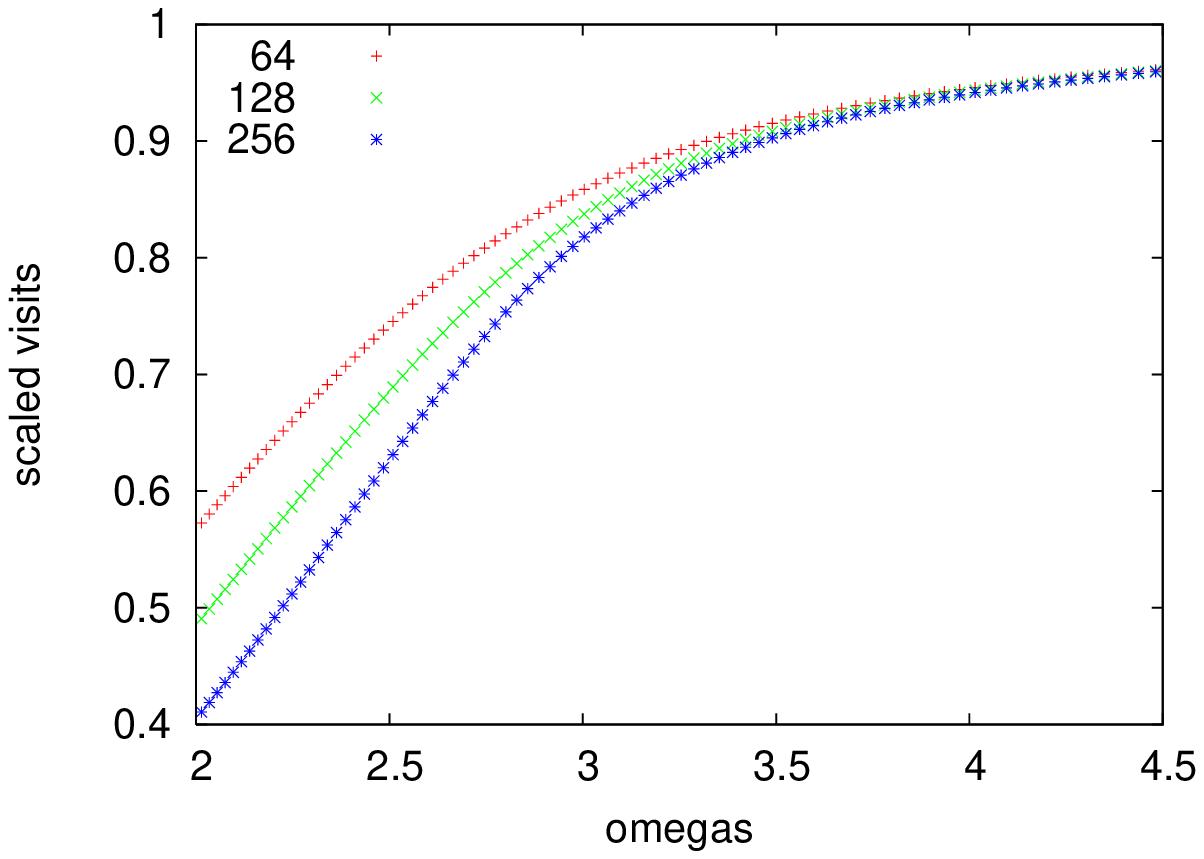}}
  \caption{Plot of the mean number of visits divided by $n^\delta$ vs
    $\omega_w$ for $\omega_p=2.0$. (a) shows the scaling in the FDG
    regime with $\delta=0$. (b) shows the scaling in the SAG regime
    with $\delta=2/3$. (c) shows the scaling in the thermodynamic
    adsorbed extended phase with $\delta=1$.}
  \label{fig comp sag fdg ob2}
\end{figure}

We now turn to considering analysis of the line $\omega_p=2.0$. We
first reinforce the analysis of Figure~\ref{fig sag exists}
and~\ref{fig fdg exists} by plotting the mean number of visits divided
by $n^\delta$ along the line $\omega_p=2.0$. We see that using
$\delta=0$ we identify the FDG regime for $\omega_w < 1$ in
Figure~\ref{fig comp sag fdg ob2}~(a). We also identify the SAG regime
by using $\delta={2/3}$ in Figure~\ref{fig comp sag fdg ob2}~(b) for
$1 \lesssim \omega_w $ and at least $\omega_w <2$, but probably for
large $\omega_w$ also. Finally the Adsorbed-Extended phase can be seen
for $\omega_w>3$ where $\delta=1$.

\begin{figure}[ht!]
  \centering
  \psfrag{omegas}{$\omega_w$}
  \includegraphics[width=7cm]{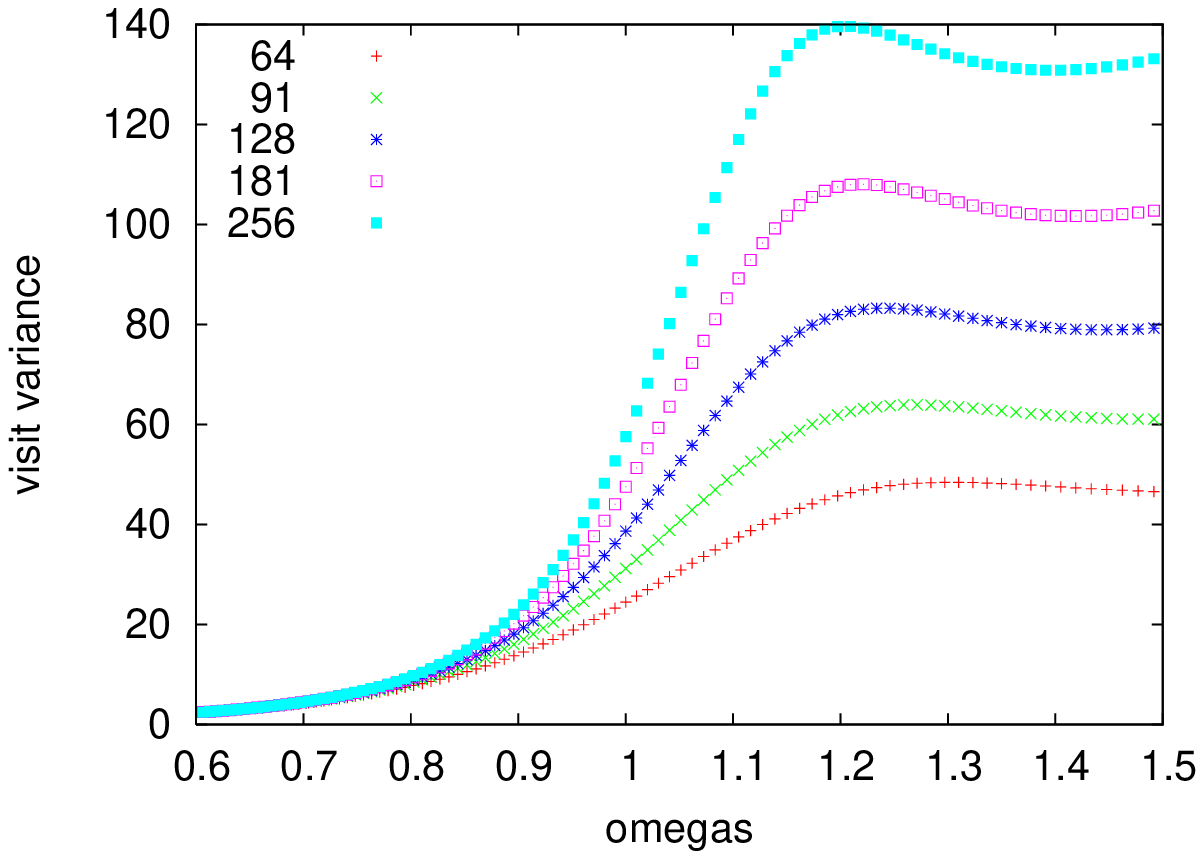}
  \caption{The variance in the number of visits vs $\omega_w$ at
    $\omega_p=2.0$ for different lengths around the SAG-FDG boundary.}
  \label{fig sagfdg boundary}
\end{figure}

Considering the variance in the number of contacts (Figure~\ref{fig
  twin peaks}), the bulk adsorption transition can be seen as a peak
developing around $\omega_w \approx 2.5$, while another smaller peak
can be identified just above $1.0$ (see Figure~\ref{fig sagfdg
  boundary}). While one might try to extrapolate the peak positions in
Figure~\ref{fig sagfdg boundary} to find the thermodynamic limit
position, we find that our data is consistent with a wide range of limits ranging
from 0.9 up to 1.15, depending on the choice of the correction exponent used to
do the extrapolation. Certainly the peaks are moving to smaller values
of $\omega_w$ as the length is increased.
\begin{figure}[pht!]
  \centering
  \psfrag{omegas}{$\omega_w$}
  \includegraphics[width=7cm]{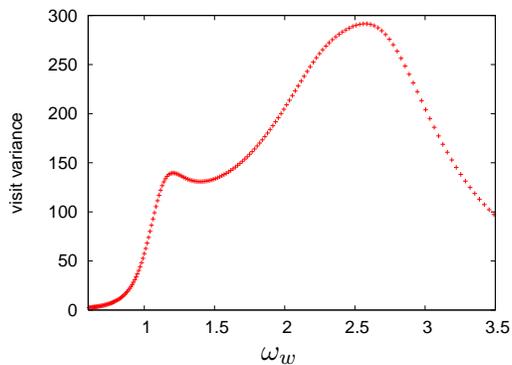}
  \caption{The variance in the number of visits vs $\omega_w$ at
    $\omega_p = 2.0$ and length $256$. One can clearly see two peaks
    indicating the possibility of two transitions.}
  \label{fig twin peaks}
\end{figure}

\begin{figure}[ht!]
  \centering
  \psfrag{omegab}{$\omega_p$}
  \includegraphics[width=7cm]{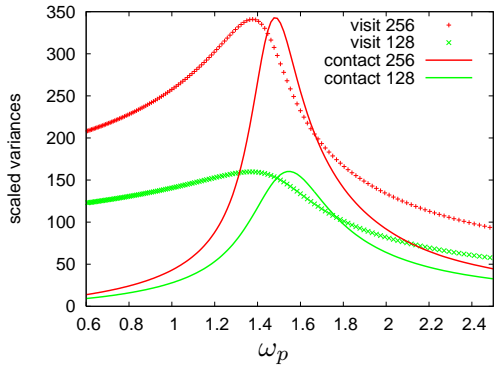}
  \caption{Variance in wall visits and bulk contacts at
    $\omega_w=1.3$ vs $\omega_p$ at lengths $128$ and $256$. Note that the peak
    in the variance of the visits is to the left of the peak of the
    variance of the number of contacts. The variance of the number of
    contacts has been scaled so that the peak height of a given length
    is approximately the same as the peak height of the variance of
    the number of visits at the same length.}
  \label{fig bulk surf var}
\end{figure}

We now return to the analysis of the line $\omega_w=1.3$.  In
Figure~\ref{fig bulk surf var}, we plot the variances in wall visits
and bulk contacts for lengths $128$ and $256$ on the same figure,
scaling the bulk variance so that the corresponding peak heights of
the wall visit variances are roughly comparable. This enables us to
make an easier comparison of the peak positions at the two lengths.
We reiterate that only one peak occurs in the variance of the wall
visits.  If indeed this peak were associated with a surface transition
occurring at a higher value of $\omega_p$ than the bulk transition,
one would expect to see the peaks of the variance of the visits
extrapolating to a higher value of $\omega_p$ than the peaks of the
contacts variances.  In both cases the peaks of the variances in wall
visits occur at smaller values of $\omega_p$ than the peaks of the
bulk contacts, and both are moving to smaller $\omega_p$ as the length
is increased.  We have investigated the scaling of the peak heights of
the variances in wall visits and bulk contacts illustrated in
Figure~\ref{fig bulk surf var}. As expected the peak height of the
(normalised) variance of the bulk contacts is consistent with a
logarithmic scaling --- an effective scaling as $(\log(n))^{2.8}$ is
seen (one expects for very large $n$ scaling of $((\log(n))^{3/11}$).
The scaling of (normalised) variance of the surface visits is
consistent with a logarithmic scaling using an effective form of
$(\log(n))^{0.5}$ (or a weak power of $n^{0.1}$). Presumably, if both
have logarithmic scaling this would be consistent with both peaks
being associated with the bulk collapse transition.  The behaviour
observed does not support the scenario in Figure~\ref{fig
  hypotheses}~(a). This suggests that there exists one phase
transition in the thermodynamic limit that gives rise to singularities
in the scaling of the contacts and the wall visits.

\begin{figure}[ht!]
   \centering \subfigure[]{
    \includegraphics[width=6cm]{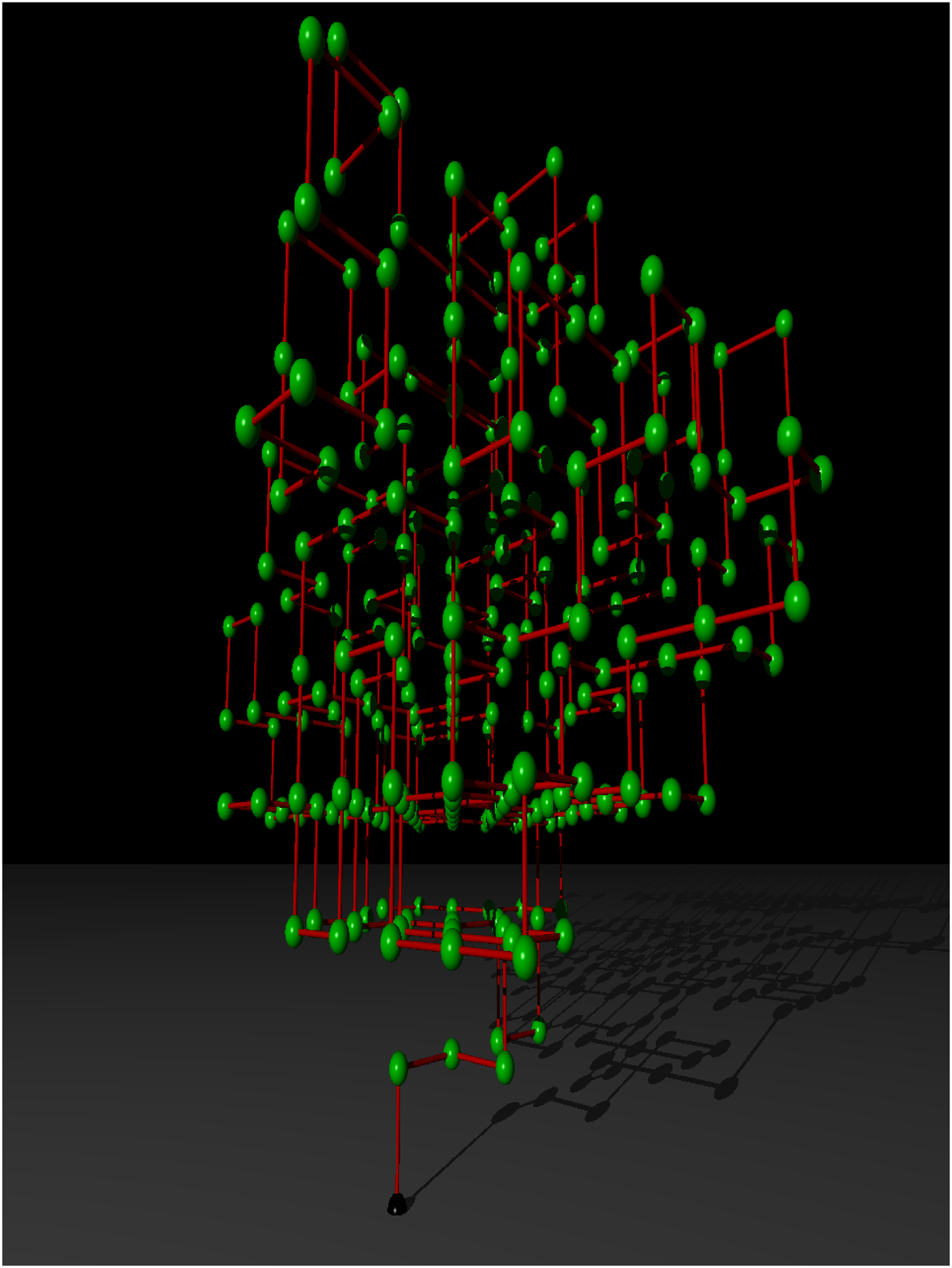}} \hspace{0.5cm}\subfigure[]{
    \includegraphics[width=6cm]{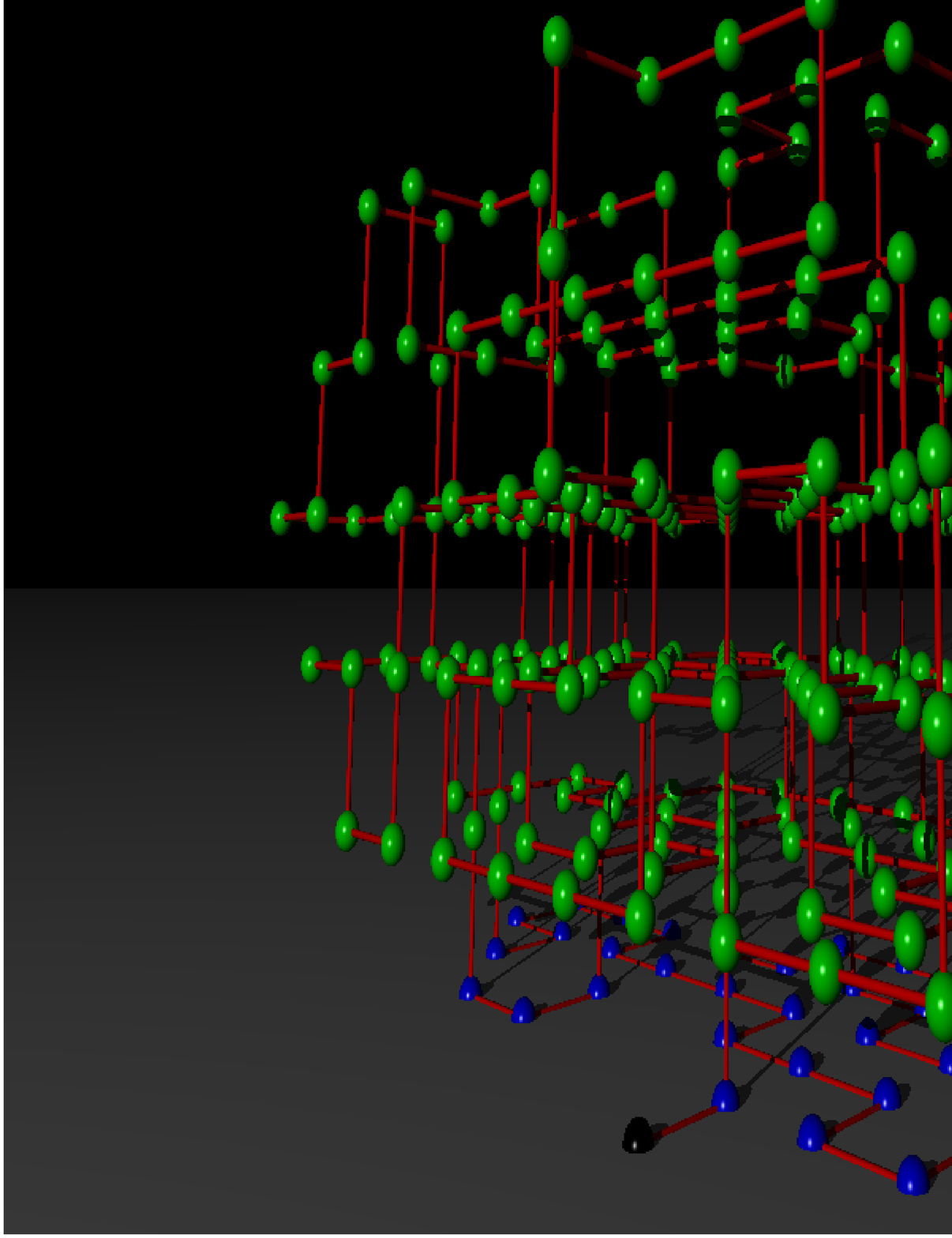}}
  \caption{Examples  of typical configurations of length $n=256$ at the points
$(\omega_p,\omega_w)=(2,0.8)$ (left) and $(2,1.3)$ (right).}
  \label{configs}
\end{figure}

Finally, we show typical examples of configurations of length $n=256$
in Figure~\ref{configs} at the points $(\omega_p,\omega_w)=(2,0.8)$
and $(2,1.3)$ which we have argued are in the FDG and SAG surface
phases respectively. One clearly sees a globule like conformation in
both examples with a dense amorphous grouping of monomers. At
$(2,0.8)$ only the monomer tethered to the wall lies in contact with
the wall while at $(2,1.3)$ a fair proportion of the monomers of the
surface of the globule lie on the wall as one would expect in a
surface-attached globule.

We conclude by summarising that while we verify the existence of the
SAG phase we find no indication of separate transitions other than
occur in the bulk phase diagram or along zero interaction boundaries.
This gives rise to a hypothesised surface phase diagram as illustrated
in Figure~\ref{fig hypotheses}~(b). 


\newpage

\section*{Acknowledgements}
The authors thank the Australian Research Council and NSERC of Canada
for financial support.


\providecommand{\newblock}{}

\end{document}